\documentclass[11pt,a4paper]{article}
\usepackage[margin=1in]{geometry}
\usepackage[T1]{fontenc}
\usepackage{lmodern}
\usepackage{microtype}
\usepackage{amsmath,amssymb,amsthm}
\usepackage{bm}
\usepackage{graphicx}
\usepackage{booktabs}
\usepackage{multirow}
\usepackage{float}
\usepackage[round]{natbib}
\usepackage{xcolor}
\usepackage{hyperref}
\hypersetup{colorlinks=true, linkcolor=blue!60!black, citecolor=blue!50!black,
            urlcolor=blue!60!black,
            pdftitle={Local Gaussian Correlation in the Tails: A Scarcity
                      Diagnostic, an Optimal Local Bandwidth, and the Limits of
                      Adaptivity},
            pdfauthor={Akash Deep, Gagan Deep},
            pdfsubject={Nonparametric dependence; local Gaussian correlation;
                        bandwidth selection; tail dependence},
            pdfkeywords={local Gaussian correlation; tail dependence; adaptive
                         bandwidth; bandwidth selection; local likelihood;
                         nonparametric dependence; copula; effective sample size;
                         AMISE; kernel estimation; financial contagion;
                         normal tempered stable}}

\graphicspath{{figures/}{../figures/}}

\newcommand{\effn}{\mathrm{eff\_n}}
\newcommand{\E}{\operatorname{\mathbb{E}}}
\newcommand{\Var}{\operatorname{Var}}

\newcommand{\Stove}{St\o{}ve}
\newcommand{\x}{\bm{x}}
\newcommand{\z}{\bm{z}}

\theoremstyle{plain}

\newtheorem{theorem}{Theorem}

\title{Local Gaussian Correlation in the Tails:\\
A Scarcity Diagnostic, an Optimal Local Bandwidth,\\
and the Limits of Adaptivity}
\author{Akash Deep\\ \small\texttt{akash.deep@ttu.edu}
  \and Gagan Deep\\ \small\texttt{gdeep@ttu.edu}}
\date{\today}

\begin{document}
\maketitle

\begin{abstract}
\noindent
Local Gaussian correlation (LGC) measures dependence \emph{locally}, making it a
natural tool for tail dependence and financial contagion, but its estimates degrade
in the joint tails, where they are most needed. Location-adaptive
bandwidths have been tried for LGC and found inferior to a single global bandwidth;
we explain why, and map the regime in which adaptivity does help. First, a
diagnostic: across heavy-tailed data-generating processes the parametric marginal
pre-transform is \emph{inert} (it changes the integrated error only in the fourth
decimal), while the binding constraint is the \emph{local effective sample size},
with the replication dispersion following a Fisher variance floor
$\mathrm{sd}\approx(1-\rho^2)/\sqrt{\effn}$. Second, theory: specializing the
Hjort--Jones local-likelihood asymptotics to the bivariate Gaussian family that LGC
fits, we derive the first \emph{location-specific AMISE-optimal bandwidth} for LGC,
$b^\star(\x)\propto[(1-\rho^2)^2/(f\beta^2)]^{1/6}n^{-1/6}$, and validate its bias
expansion directly (bias $\propto b^2\beta$, $R^2\approx0.9$, slope-to-$\beta$
correlation $0.80$). Third, a regime map: a Monte Carlo across dependence strengths
shows the adaptive rule beats the global plug-in \emph{only at moderate dependence
with curved surfaces}. At weak dependence there is no curvature to exploit; at
strong dependence finite-sample bias from the steep surface dominates, and adaptivity
performs substantially worse, with an error that grows in the sample size. This
explains the field's experience that global bandwidths are hard to beat, and locates
the exception. Fourth, application: on volatility-filtered equity returns the
adaptive estimator yields more stable tail-dependence surfaces under resampling. The
message is cautionary: the binding constraint on tail LGC is data scarcity, not
bandwidth placement, and no bandwidth, however optimal, can recover information the
data do not contain.
\end{abstract}



\section{Introduction}
\label{sec:intro}

Dependence between financial assets is not constant across their range: equities
that are nearly uncorrelated in calm markets crash together in a crisis. Capturing
this requires a \emph{local} measure of dependence. Local Gaussian correlation
\citep{tjostheim2013local, berentsen2014recognizing} provides one: at each point of
the sample space it fits a bivariate Gaussian by kernel-weighted local likelihood
and reads off the correlation $\rho(\x)$, yielding a full dependence \emph{surface}
that reduces to the global correlation only when the data are jointly Gaussian. LGC
has been used to re-examine financial contagion \citep{stove2014using}, allocate
portfolios under asymmetric dependence \citep{sleire2022portfolio}, and detect
regime changes \citep{gundersen2024local}.

The difficulty is that LGC is least reliable where it is most valuable. A local
estimator needs observations in the local neighbourhood, and the joint tails are by
definition sparse. The natural lever is the bandwidth, and one might hope a
location-adaptive bandwidth that widens in the tails would help. It largely does
not: an adaptive bandwidth was tried for the locally Gaussian density estimator and
found inferior to a single global selector \citep{otneim2017locally}, and global
bandwidths remain the default. This paper explains that experience and bounds it. We
identify what governs tail error, derive the AMISE-optimal local bandwidth
for LGC, and map the narrow regime in which location-adaptivity helps and the wider
regimes in which it does not.

Our contributions:
\begin{enumerate}
\setlength{\itemsep}{1pt}
\item \textbf{The marginal transform is inert} (\S\ref{sec:inert}). Fitting a
  heavy-tailed parametric marginal before LGC, a natural-seeming fix for fat-tailed
  data, changes the integrated error only in the fourth decimal. The lever is not
  the marginal.
\item \textbf{A scarcity diagnostic} (\S\ref{sec:scarcity}). Tail scarcity has been
  noted informally \citep{jordanger2022nonlinear}; we make it precise. LGC error is
  organised by the local effective sample size $\effn(\x)$, not by the estimator
  variant, and its dispersion obeys a Fisher variance floor $\mathrm{sd}\approx
  (1-\rho^2)/\sqrt{\effn}$. This identifies the bandwidth as the operative lever.
\item \textbf{An AMISE-optimal local bandwidth} (\S\ref{sec:theory}). We derive
  $b^\star(\x)\propto[(1-\rho^2)^2/(f\beta^2)]^{1/6}n^{-1/6}$ by balancing the
  local-likelihood bias against this variance floor. To our knowledge this is the
  first \emph{AMISE-optimal, location-specific} bandwidth derived for the LGC
  estimator; we validate its bias expansion directly. Prior LGC bandwidths are
  global, and the one prior adaptive attempt was an ad hoc nearest-neighbour device
  found inferior to the global selector \citep{otneim2017locally}.
\item \textbf{A regime map for adaptivity} (\S\ref{sec:mc}). A Monte Carlo across
  dependence strengths shows the adaptive rule beats the global plug-in \emph{only at
  moderate dependence with curved surfaces}. At weak dependence the surface is flat
  and redistribution only adds variance; at strong dependence the surface is steep
  and finite-sample bias dominates, making adaptivity substantially worse with an
  error that grows in the sample size; an oracle-scale experiment shows this
  failure is intrinsic to the pointwise-optimal shape, not to our normalisation.
  This explains why global bandwidths have been hard to beat, and locates the
  exception. On real equity returns
  (\S\ref{sec:real_data}) the adaptive estimator is more stable under resampling, a
  use that does not require beating global on accuracy.
\end{enumerate}

The common thread: tail LGC is bounded by data scarcity, which a bandwidth can
redistribute but not overcome.

\section{Background: the LGC estimator and its bandwidth}
\label{sec:background}

Let $\z=(z_1,z_2)$ have joint density $f$ on $\mathbb R^2$. LGC fits, at a point
$\x$, a bivariate Gaussian $\psi(\cdot\,;\bm\theta)$ with
$\bm\theta=(\mu_1,\mu_2,\sigma_1,\sigma_2,\rho)$ by maximising the kernel-weighted
local log-likelihood of \citet{hjort1996locally},
\begin{equation}
L_n(\bm\theta;\x)=\frac1n\sum_{i=1}^n K_b(\bm Z_i-\x)\log\psi(\bm Z_i;\bm\theta)
   -\int K_b(\bm u-\x)\,\psi(\bm u;\bm\theta)\,d\bm u,
\label{eq:llik}
\end{equation}
with a product kernel $K_b(\bm u)=b^{-2}K(u_1/b)K(u_2/b)$. The maximiser
$\hat{\bm\theta}(\x)$ defines the local Gaussian correlation $\hat\rho(\x)$. In
practice each margin is first mapped to standard normality by the empirical
normal-score transform, so the estimator operates on Gaussian pseudo-observations
\citep{berentsen2014recognizing}; \S\ref{sec:inert} shows this choice of marginal
transform is immaterial.

\paragraph{Bandwidth.} Bandwidth selection for LGC has, to date, been
\emph{global}. Two choices dominate: a global likelihood cross-validation
\citep{berentsen2014departures, otneim2017locally} and a fast plug-in
$b_0=1.75\,n^{-1/6}$ whose rate follows from the local-likelihood asymptotics and
whose constant is fixed empirically \citep{tjostheim2013local}; both are the only
options in the standard software \citep{berentsen2014localgauss} and underpin the
methodology of the reference monograph \citep{tjostheim2022statistical}. A point-dependent
$k$-nearest-neighbour bandwidth was tried in the locally Gaussian density setting but
found inferior to the global selector and not pursued \citep{otneim2017locally}, and
the data-scarcity of distributional tails has been noted only informally, with ``use
a larger bandwidth'' offered as the remedy \citep{jordanger2022nonlinear}. To our
knowledge, no prior work derives a bias--variance-balanced, location-specific
bandwidth for the LGC estimator itself; that is the gap this paper fills. We write the \emph{local effective
sample size}
\begin{equation}
\effn(\x)=\sum_i K_b(\bm Z_i-\x)\Big/\max_i K_b(\cdot)
   \;\approx\;2\pi b^2 n\, f(\x)\quad(\text{Gaussian }K),
\label{eq:effn}
\end{equation}
the number of observations that contribute meaningfully to the local fit; it is the
quantity the rest of the paper turns on.

\paragraph{Monte Carlo design.} Unless stated otherwise, simulations use four copulas
(Gaussian, Clayton, $t_4$, Gumbel) parameterised to Kendall's $\tau=0.5$, pushed
through symmetric unit-variance normal tempered stable marginals
\citep{rachev2011financial}, with $R=50$
replications at $n\in\{500,1000,2500,5000\}$ (the dependence sweep of
\S\ref{sec:regime} varies $\tau$ and uses $R=25$). ``Ground truth'' is LGC on an auxiliary
$n_{\mathrm{gt}}=20{,}000$ sample, evaluated on an $11\times11$ grid over
$[-3,3]^2$ plus five diagonal quantile points $\z_q=(\Phi^{-1}(q),\Phi^{-1}(q))$,
$q\in\{.01,.05,.5,.95,.99\}$. Accuracy is the density-weighted integrated squared
error (ISE) over the grid; tail behaviour is read at the quantile points.

\section{The marginal transform is inert}
\label{sec:inert}

Heavy-tailed marginal models are standard in financial econometrics
\citep{barndorff1997normal, cont2004financial, rachev2011financial,
deep2025probability}, so a natural hypothesis is that fitting a parametric
heavy-tailed marginal before estimating LGC, rather than using the empirical
transform, should sharpen the tail estimates. We test this with three estimators differing \emph{only}
in the marginal step: canonical (empirical CDF), an oracle that transforms through
the true marginal CDF, and a version that fits the marginal by maximum likelihood.
Because the oracle carries no marginal-estimation error by construction, the
canonical--oracle gap isolates the transform's contribution.

\begin{figure}[H]
\centering
\includegraphics[width=\textwidth]{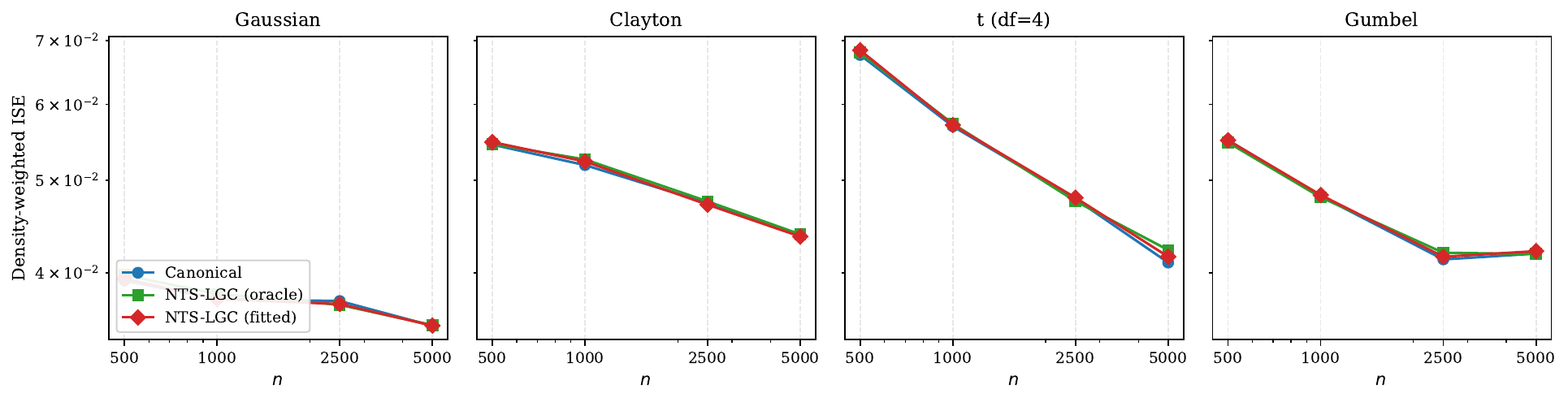}
\caption{Density-weighted ISE versus $n$ (log--log), by copula. The three
estimators, which differ only in the marginal transform, sit on top of one another
in every panel.}
\label{fig:ise}
\end{figure}

That gap is negligible. At $n=5000$ the three estimators agree to the fourth decimal
of density-weighted ISE in every copula (Table~\ref{tab:inert},
Figure~\ref{fig:ise}), while the tail sampling error is one to two orders of
magnitude larger. This quantifies, under heavy tempered-stable tails, the
marginal-invariance already implicit in the standard practice of estimating LGC on
normal-score pseudo-observations \citep{berentsen2014recognizing}: local dependence
is a copula property, so the marginal transform contributes only estimation noise.
Nor does asymmetry change this: repeating the comparison with a skewed NTS marginal
($\beta=-0.6$, skewness $\approx-0.84$; $n=2500$, $R=25$) leaves the three estimators
within $0.0003$--$0.0014$ of one another in ISE across all four copulas; the
conclusion is unchanged. The marginal is not the lever; we fix the canonical
empirical transform hereafter and ask what is.

\begin{table}[H]
\centering\small
\begin{tabular}{lccc}
\toprule
Copula & Canonical & LGC (oracle marginal) & LGC (fitted marginal) \\
\midrule
Gaussian & 0.0352 (0.0067) & 0.0353 (0.0068) & 0.0352 (0.0066) \\
Clayton  & 0.0438 (0.0070) & 0.0439 (0.0074) & 0.0437 (0.0070) \\
$t_4$    & 0.0410 (0.0082) & 0.0423 (0.0093) & 0.0416 (0.0087) \\
Gumbel   & 0.0419 (0.0071) & 0.0419 (0.0076) & 0.0421 (0.0074) \\
\bottomrule
\end{tabular}
\caption{Density-weighted ISE at $n=5000$, with the standard deviation across the
$R=50$ replications in parentheses. Within-row differences sit in the fourth
decimal, an order of magnitude below the replication SD: the marginal transform is
inert.}
\label{tab:inert}
\end{table}

\section{The binding constraint is local scarcity}
\label{sec:scarcity}

\begin{figure}[H]
\centering
\includegraphics[width=\textwidth]{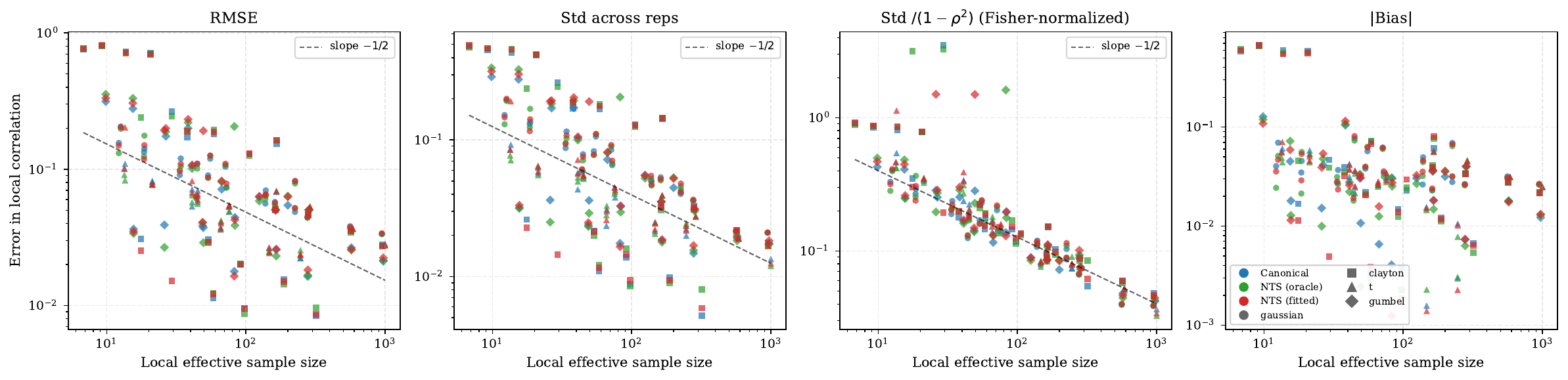}
\caption{LGC error against local effective sample size, pooled over all cells. Left
to right: RMSE, replication sd, Fisher-normalised sd $\mathrm{sd}/(1-\rho^2)$
(tightest collapse, slope $-1/2$ guide), and $|$bias$|$. Colour marks the estimator
(they overlap), shape the copula.}
\label{fig:scarcity}
\end{figure}

If neither the estimator variant nor the marginal separates the cells, what governs
the error? The local effective sample size~\eqref{eq:effn}. That distributional tails
are data-poor for LGC has been remarked before, informally, with a larger bandwidth
suggested as the countermeasure \citep{jordanger2022nonlinear}; here we make the
observation quantitative. Two facts stand out (Figure~\ref{fig:scarcity}).

First, \textbf{error is organised by $\effn$, not by the estimator}: within each
cell the RMSE spread across the three estimators of \S\ref{sec:inert} has median
$0.005$, while $\effn$ itself falls from $\sim\!160$--$1000$ at the centre to as few
as $\sim\!7$ in the tails.

Second, \textbf{the dispersion is a Fisher variance floor}. The replication standard
deviation follows $\mathrm{sd}\approx(1-\rho^2)/\sqrt{\effn}$; regressing
$\log[\mathrm{sd}/(1-\rho^2)]$ on $\log\effn$ gives slope $-0.61$ with $R^2=0.76$.
This matches the asymptotic variance of a Gaussian-correlation maximum
likelihood estimator from $m$ observations, $\Var(\hat\rho)\approx(1-\rho^2)^2/m$,
with $m=\effn$ the local count. The fitted slope is somewhat steeper than the
theoretical $-1/2$; the excess is consistent with the sparsest cells, where the
floor is least clean and bias contaminates the dispersion, pulling the pooled
slope down. Most tail error is therefore variance, and variance is bought back by
raising $\effn$.

Bias dominates only at the sharply-structured sparse corners; the clearest is
Clayton at $q=0.99$ (true local $\rho\approx-0.68$, $\effn\approx7$), where
$\mathrm{RMSE}\approx0.76$ but $|\mathrm{bias}|\approx0.60\gtrsim\mathrm{sd}\approx
0.48$. This is the regime no bandwidth can rescue, and we return to it below.

The consequence for design: a bandwidth that raises $\effn$ in the tails
has real headroom against the variance floor, but widening blindly trades variance
for bias wherever the surface is not locally constant. The bandwidth, not the
marginal, is the lever, and it must widen with sparsity while respecting curvature.

\section{An AMISE-optimal local bandwidth}
\label{sec:theory}

\subsection{Local bias and variance}

The LGC estimator is an instance of the \citet{hjort1996locally} local-likelihood
estimator, so its bias and variance follow from that theory specialized to the
five-parameter Gaussian family. The two ingredients exist separately in the LGC
literature: the leading $O(b^2)$ bias of the local Gaussian likelihood estimator was
derived by \citet{otneim2013bias}, and the asymptotic covariance of the
local-correlation maximum likelihood estimator was given by \citet{otneim2021partial}.
Our contribution here is to assemble them into the first explicit local-AMISE balance
for $\rho(\x)$. Under standard conditions ($f$ smooth and bounded away from $0$ near
$\x$; $b\to0$, $nb^2\to\infty$), the correlation component satisfies
\begin{align}
\Var[\hat\rho(\x)] &= \frac{R(K)^2}{n\,b^2 f(\x)}\,\big(1-\rho(\x)^2\big)^2
   + o\!\big(\tfrac1{nb^2}\big),
\label{eq:var}\\
\E[\hat\rho(\x)]-\rho(\x) &= \tfrac12\,\mu_2(K)\,b^2\,\beta(\x) + o(b^2),
\label{eq:bias}
\end{align}
with kernel constants $\mu_2(K)=\int u^2K$, $R(K)=\int K^2$. The variance
factor $(1-\rho^2)^2$ is the per-observation asymptotic variance of the
Gaussian-correlation MLE; dividing by the local count~\eqref{eq:effn} reproduces the
empirical Fisher floor of \S\ref{sec:scarcity}, tying the diagnostic to the theory.
The bias functional, to leading order,
\begin{equation}
\beta(\x)=\Delta\rho(\x)+2\,\nabla\log f(\x)^{\!\top}\nabla\rho(\x)+r(\x),
\label{eq:beta}
\end{equation}
combines the Laplacian of the local-correlation surface (the curvature term) with a
random-design density-drift term and a model-misfit remainder $r$ that vanishes
where $f$ is locally Gaussian. \S\ref{sec:biasval} validates~\eqref{eq:bias} directly;
the exact closed-form constant in~\eqref{eq:beta}, a full specialization of
Hjort--Jones to the Gaussian family, is deferred to future work and is not needed
for the bandwidth, which is invariant to it (\S\ref{sec:method}).

\subsection{The optimal bandwidth}

The local asymptotic mean squared error is, from \eqref{eq:var}--\eqref{eq:bias},
\begin{equation}
\mathrm{AMSE}(\x;b)=\tfrac14\mu_2(K)^2\,b^4\,\beta(\x)^2
   +\frac{R(K)^2\,(1-\rho(\x)^2)^2}{n\,b^2 f(\x)} .
\label{eq:amse}
\end{equation}
Minimising over $b$ yields the location-specific optimum.

\begin{theorem}[AMISE-optimal LGC bandwidth]
\label{thm:bw}
The bandwidth minimising the local AMSE~\eqref{eq:amse} is
\begin{equation}
b^\star(\x)=C_K\left[\frac{\big(1-\rho(\x)^2\big)^2}{f(\x)\,\beta(\x)^2}\right]^{1/6}
   n^{-1/6},
\qquad C_K=\left(\frac{2R(K)^2}{\mu_2(K)^2}\right)^{1/6}.
\label{eq:bstar}
\end{equation}
\end{theorem}
The optimal rate $n^{-1/6}$ matches the canonical global plug-in; what is new is the
location-specific \emph{constant}. The bandwidth widens in sparse regions
($\propto f^{-1/6}$), narrows where the surface is structured
($\propto\beta^{-1/3}$, the curvature penalty), and \emph{narrows} toward the
dependence extremes ($\propto(1-\rho^2)^{1/3}$): where $|\rho|\to1$ the
per-observation variance $(1-\rho^2)^2$ collapses, so less local averaging is needed.
\S\ref{sec:mc} shows the cost of this last property: the over-narrowing, and the
compensating over-widening it forces elsewhere, is what makes the rule misbehave
when the whole surface is strongly dependent.

\subsection{Correction of a borrowed heuristic}

A natural way to make the LGC bandwidth adaptive is to import the
\citet{brockmann1993locally} curvature/variance balance from one-dimensional kernel
regression, part of a long variable-bandwidth tradition in density and regression
smoothing \citep{breiman1977variable, abramson1982bandwidth, loader1996local},
augmented with the random-design density widening of \citet{fan1996local}; doing so
gives $b_{\text{heur}}\propto(1-\rho^2)^{1/6}f^{-1/6}\kappa^{-1/3}$ with
$\kappa=\lVert\nabla^2\rho\rVert$. Theorem~\ref{thm:bw}
shows two corrections are required. The dependence factor is $(1-\rho^2)^{1/3}$, not
$(1-\rho^2)^{1/6}$: the borrowed rule takes a square root too few, because the LGC
variance scales as $(1-\rho^2)^2$ rather than $(1-\rho^2)$. And the curvature is the
\emph{signed} functional $\beta$ of~\eqref{eq:beta}, with its density-drift term,
rather than the Frobenius Hessian norm $\kappa$. The first correction makes the rule
narrow more aggressively at high $|\rho|$. As \S\ref{sec:mc} shows, this helps at
moderate dependence, where the corrected rule improves on the heuristic, and hurts
at strong dependence, where the same aggressiveness destabilises it.

\subsection{Numerical validation of the bias expansion}
\label{sec:biasval}

Equation~\eqref{eq:bias} is the load-bearing assumption behind
Theorem~\ref{thm:bw}, so we test it directly. For curved-surface copulas (Clayton,
Gumbel) we sweep a ladder of bandwidths, averaging $\hat\rho(\x;b)$ over $120$
replications at each $b$ ($n=4000$), and at each grid point regress the mean estimate
on $b^2$; its slope estimates $\tfrac12\mu_2\beta(\x)$ without reference to the
unknown $\rho_0$ (the intercept).

\begin{figure}[H]
\centering
\includegraphics[width=\textwidth]{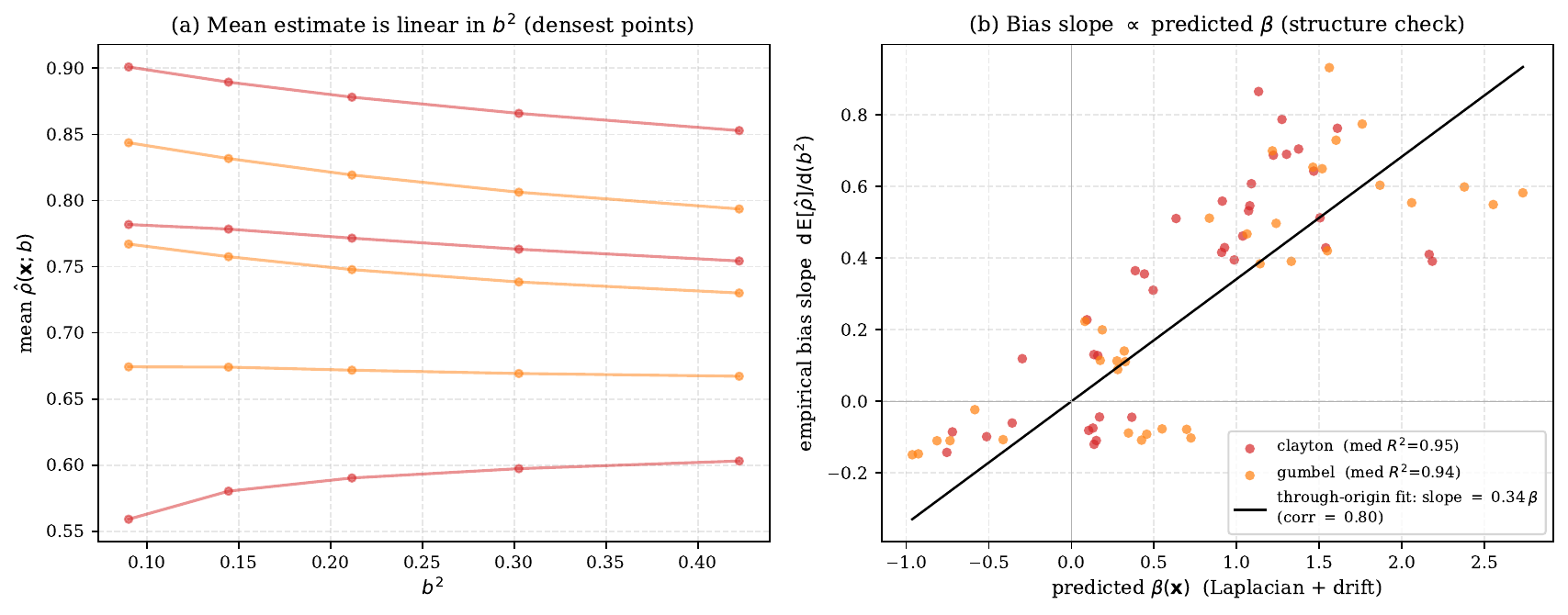}
\caption{Validation of the bias expansion~\eqref{eq:bias}. (a) The mean estimate is
linear in $b^2$ at the densest points. (b) The empirical bias slope
$\mathrm{d}\,\E[\hat\rho]/\mathrm{d}(b^2)$ is proportional to the predicted
$\beta(\x)$ (correlation $0.80$, through-origin fit shown), confirming the bias
structure.}
\label{fig:biasval}
\end{figure}

Both halves hold (Figure~\ref{fig:biasval}). The \emph{rate}: the per-point fits are
strongly linear in $b^2$ (median $R^2=0.93$ and $0.90$ for the two copulas). The
\emph{structure}: the empirical bias slope is proportional to the predicted
$\beta(\x)$ of~\eqref{eq:beta}, correlation $0.80$ over $73$ points, sign-correct
through the origin. The fitted proportionality constant ($0.34$) sits below the
nominal $\mu_2/2$, as expected: the exact Hjort--Jones constant carries projection
factors beyond $\mu_2/2$, and estimating $\beta$ by finite differences on a pilot
grid attenuates any such regression slope toward zero. The \emph{form}
underlying~\eqref{eq:bstar}, which is all the bandwidth requires, is confirmed.

\section{The adaptive estimator}
\label{sec:method}

Theorem~\ref{thm:bw} is operationalised by plug-in. All three location functionals
come from a single global-bandwidth pilot fit: the density $\hat f$ from the pilot
$\effn$ via \eqref{eq:effn}; $1-\hat\rho^2$ from the pilot surface; and
$\hat\beta=\widehat{\Delta\rho}+2\,\widehat{\nabla\log f}^{\top}\widehat{\nabla\rho}$
by finite differences. Two design choices keep the comparison fair and the procedure
stable.

\paragraph{Budget neutrality.} We normalise the per-point constant so that its
density-weighted geometric mean equals the global plug-in constant. The adaptive rule
then \emph{redistributes} a fixed smoothing budget (wider tails, narrower
centre) rather than globally inflating it, so any gain is attributable to placement,
not to more smoothing. This also makes $b^\star$ invariant to the unknown overall
constant $C_K$ and to the closed-form constant in $\beta$.

\paragraph{Curvature stabilisation.} Because $\beta$ sits in a denominator, a
near-empty, sharply-structured tail could otherwise produce a runaway window. We
smooth $\beta$ at a coarse scale, floor it away from zero, and cap the per-point
bandwidth to $[0.5,3]\times$ the global value; these are the stabilisers
\citet{brockmann1993locally} show are necessary for an estimated local bandwidth.

\section{When does adaptivity help? A Monte Carlo}
\label{sec:mc}

We compare the global plug-in against the adaptive rules on shared samples (a paired
design that removes sampling noise from the differences). Two questions: at the
conventional moderate dependence, whether and where adaptivity helps
(\S\ref{sec:moderate}); and across dependence strength, the regime in which it helps
at all (\S\ref{sec:regime}).

\subsection{At moderate dependence ($\tau=0.5$)}
\label{sec:moderate}

\begin{figure}[H]
\centering
\includegraphics[width=\textwidth]{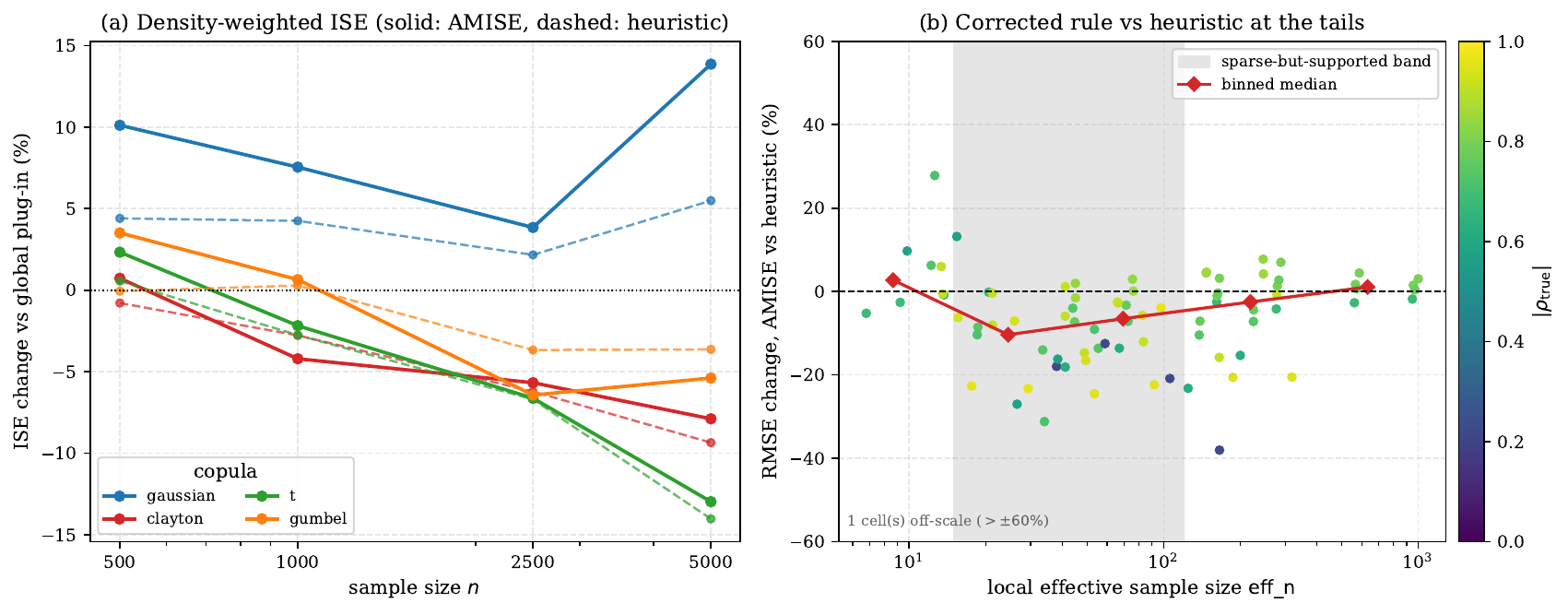}
\caption{(a) Density-weighted ISE change versus the global plug-in, AMISE (solid) and
heuristic (dashed): both beat global on curved-surface copulas and track each other
in the bulk. (b) RMSE change of the corrected rule versus the heuristic at the
quantile points, against local $\effn$, coloured by $|\rho_{\mathrm{true}}|$: the
binned median dips below zero in the shaded sparse-but-supported band and returns to
zero at both extremes. Both panels are at $\tau=0.5$.}
\label{fig:adaptive}
\end{figure}

At Kendall's $\tau=0.5$ the adaptive rules improve the density-weighted ISE on the copulas whose surface has
curvature to exploit (Clayton, $t_4$, Gumbel), the gain growing with $n$ to
$13$--$14\%$ at $n=5000$, while costing $2$--$14\%$ on the near-flat Gaussian
surface, where redistribution can only add variance (Figure~\ref{fig:adaptive}a). A
second effect was not designed for: the adaptive estimator has the smaller
replication standard deviation in $11$ of the $16$ (copula, $n$) cells, and in
\emph{every} cell with $n\ge2500$. Once the pilot fields are well estimated, it
stabilises the surface across samples.
Against local $\effn$, the picture is the one \S\ref{sec:scarcity} predicts: adaptive
is taxed below a few dozen effective observations (an estimated local bandwidth costs
$O_p(n^{-1/2})$ variability against $O_p(n^{-1})$ for a global one) and reaches
parity-or-better above.

\paragraph{Corrected rule versus the heuristic.}
Still at $\tau=0.5$, the derived rule also improves on the borrowed heuristic, and
where the theory predicts: the corrected $(1-\rho^2)^{1/3}$ exponent helps most in the
strongly-dependent, sparse tail (Table~\ref{tab:gate2},
Figure~\ref{fig:adaptive}b). Pooling the quantile-point cells by local $\effn$, the
corrected rule improves on the heuristic by a median $10\%$ in the band
$\effn\in[15,40)$ ($93\%$ of cells) and $6.5\%$ in $[40,120)$ ($81\%$ of cells),
while matching it in the dense bulk ($\effn\ge120$) and at the hopeless deepest tail
($\effn<15$). Restricted to strongly-dependent sparse cells
($|\rho|\ge0.5$, $\effn\le40$), the corrected rule beats the heuristic on $77\%$ of
cells, median $-5.7\%$. The effect is modest but predicted and consistent. At the
single worst cell (Clayton $q=0.99$, $\effn\approx7$) all three estimators are tied;
this is the bias-dominated regime no bandwidth can cure.

\begin{table}[H]
\centering\small
\begin{tabular}{lcccc}
\toprule
$\effn$ band & regime & median RMSE change & cells improved \\
\midrule
$[0,15)$   & hopeless deep tail   & $+2.7\%$  & $50\%$ \\
$[15,40)$  & sparse, supported    & $-10.4\%$ & $93\%$ \\
$[40,120)$ & moderately sparse    & $-6.5\%$  & $81\%$ \\
$[120,\infty)$ & dense bulk       & $-0.8\%$  & $55\%$ \\
\bottomrule
\end{tabular}
\caption{Corrected AMISE rule versus the borrowed heuristic at $\tau=0.5$: pointwise
RMSE change by local effective sample size (negative favours the corrected rule). The
gain is concentrated in the sparse-but-supported tail, as
Theorem~\ref{thm:bw} predicts.}
\label{tab:gate2}
\end{table}

\subsection{Across dependence strength: the regime of applicability}
\label{sec:regime}

\begin{figure}[H]
\centering
\includegraphics[width=\textwidth]{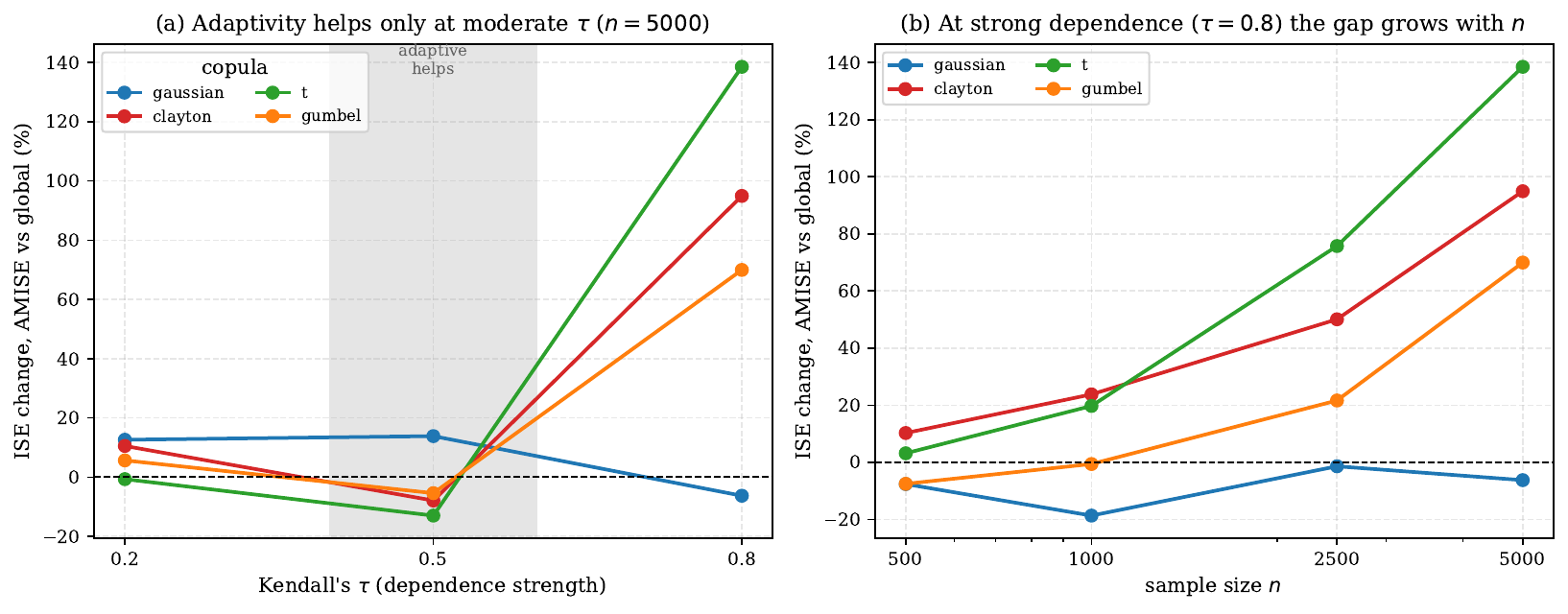}
\caption{The adaptive rule helps only at moderate dependence. (a) ISE change versus
the global plug-in against Kendall's $\tau$ at $n=5000$ (positive is worse): the
curved-copula curves dip meaningfully below zero only near $\tau=0.5$ (shaded) and
rise steeply at strong dependence; the near-flat Gaussian is the exception,
essentially flat across $\tau$ with a mild benefit at $\tau=0.8$. (b) At $\tau=0.8$
the deficit grows with $n$ for the curved copulas (a bias signature), while the
Gaussian is again the exception.}
\label{fig:regime}
\end{figure}

The moderate-dependence picture does not generalise. We sweep Kendall's
$\tau\in\{0.2,0.5,0.8\}$, comparing the global plug-in with the derived AMISE rule on
shared samples ($25$ replications per cell at $\tau\in\{0.2,0.8\}$; the $\tau=0.5$
column reuses the run above). The adaptive advantage is confined to a band
(Figure~\ref{fig:regime}):

\begin{itemize}
\setlength{\itemsep}{1pt}
\item \textbf{Weak dependence ($\tau=0.2$):} adaptive is \emph{worse} than global or
  at best at parity ($+6$ to $+13\%$ ISE at $n=5000$ for Gaussian, Clayton and
  Gumbel; $-0.7\%$ for $t_4$; $+14$ to $+18\%$ for all four copulas at $n=500$). The
  surface is nearly flat, so there is little curvature to exploit and redistribution
  mostly adds variance: the Gaussian story of \S\ref{sec:moderate}, now for
  essentially all copulas.
\item \textbf{Moderate dependence ($\tau=0.5$):} adaptive wins on the curved copulas,
  the only regime in which it does.
\item \textbf{Strong dependence ($\tau=0.8$):} on the curved copulas adaptive is
  \emph{markedly worse} ($+95\%$ for Clayton, $+139\%$ for $t_4$, $+70\%$ for Gumbel
  at $n=5000$; the near-flat Gaussian is the lone exception at $-6\%$), and the gap
  \emph{grows with $n$} (Figure~\ref{fig:regime}b), the signature of bias rather
  than variance.
\end{itemize}

\paragraph{The mechanism of the strong-dependence failure.}
The causal chain has three links. First, on a strongly dependent surface the
optimal rule~\eqref{eq:bstar} \emph{narrows} sharply along the dense,
high-$|\rho|$ ridge: the per-observation variance $(1-\rho^2)^2$ is small there,
so the pointwise bias--variance trade favours little smoothing. Second, the
narrowing collides with the smoothing budget: because the normalisation pins the
(density-weighted geometric) mean bandwidth, the budget freed on the ridge must be
spent elsewhere, and it lands on the steep mid-region, driving bandwidths into the
upper cap ($36\%$ of grid points at $\tau=0.8$, versus none in the dense bulk at
$\tau=0.5$). Third, the over-widened windows average across genuinely varying
correlation, producing bias, and bias does not shrink as $n$ grows; hence the
deficit in Figure~\ref{fig:regime}b widens with the sample size. Three checks
localise the fault. A region-by-region decomposition puts the loss in the dense
centre and mid-range (each $\approx2.9\times$ the global ISE), not the sparse
corners. Tripling the pilot-grid resolution for $\beta$ barely moves it
($2.4\times\to2.2\times$ global), ruling out curvature estimation as the cause.
And the failure is intrinsic to the \emph{shape} of~\eqref{eq:bstar}, not to the
budget-neutral normalisation: keeping the AMISE shape, discarding budget
neutrality entirely, and sweeping a global multiplier over the shape (an oracle
scale, which upper-bounds any data-driven normalisation), the best multiplier
still loses to the global plug-in by $+42\%$ (Clayton) and $+51\%$ ($t_4$) at
$\tau=0.8$, $n=2500$ on paired samples. Freeing the scale narrows the
budget-neutral deficit (for $t_4$, ISE $0.149\to0.128$ against global $0.085$)
but closes nowhere near the gap: on a steep, strongly dependent surface the
pointwise-optimal allocation itself missmooths, and no choice of overall scale
repairs it.

This is the mechanism behind the field's experience that the global bandwidth is
hard to beat for LGC \citep{otneim2017locally}: outside a moderate-dependence,
curved-surface band a single global bandwidth is preferable, and the strong-dependence
deficit is large. The scarcity diagnostic of \S\ref{sec:scarcity} explains why: tail
error is variance-floored, and only where a curved surface permits trading bias for
variance at adequate $\effn$ can a location-varying bandwidth improve the trade.

\section{Real-Data Application}
\label{sec:real_data}

The Monte Carlo isolates the adaptive bandwidth's behaviour against a known
truth. On real data no such truth exists, so we are deliberate about what can be
claimed. Two things can be shown without a ground-truth surface: that the global
and adaptive estimators resolve the tail dependence \emph{differently}, in a way
consistent with the simulation; and, quantitatively, that the adaptive surface is
\emph{more stable} under resampling. The second is the real-data analogue of the
Monte Carlo finding that at moderate-to-large samples the adaptive estimator
carries the smaller replication standard deviation (\S\ref{sec:moderate}), and it
is the headline here because it needs no truth. We do not claim a real-data accuracy gain, which would not be
identifiable.

\subsection{Data and volatility filtering}

We take daily adjusted closes for two pairs with distinct dependence geometry:
SPY versus long-dated Treasuries (TLT), the sign-varying flight-to-quality
relationship; and SPY versus emerging-market equity (EEM), strong positive
dependence with lower-tail crisis contagion, the canonical setting of the local
Gaussian contagion literature \citep{stove2014using}. The sample runs
2005--2024 ($n=5030$ common trading days after filtering), spanning the 2008 and
2020 crises so the joint tails are populated.

Daily returns carry volatility clustering; LGC on raw returns would partly
measure common volatility rather than dependence. Following the contagion
literature we pre-whiten each series with an AR(1)--GARCH(1,1) model with
Student-$t$ innovations and estimate LGC on the standardised residuals
(Table~\ref{tab:garch}). The Ljung--Box statistics confirm the filter removes
both serial correlation and volatility clustering, while the residuals retain the
heavy tails and asymmetry that carry the dependence of interest. The marginal
step is then the canonical empirical normal-score transform; Result~1 (the
parametric transform is inert) is what licenses this simpler choice over a
refitted NTS margin.

\begin{table}[H]
\centering\small
\begin{tabular}{lcccccc}
\toprule
Series & $\alpha_1$ & $\beta_1$ & $\nu$ & resid.\ skew & resid.\ ex-kurt
  & LB$^2$ $p$ \\
\midrule
SPY & 0.141 & 0.854 & 5.6 & $-0.71$ & $2.56$ & 0.41 \\
TLT & 0.054 & 0.941 & 15.5 & $-0.09$ & $0.76$ & 0.13 \\
EEM & 0.097 & 0.888 & 9.1 & $-0.37$ & $1.58$ & 0.06 \\
\bottomrule
\end{tabular}
\caption{AR(1)--GARCH(1,1)-$t$ volatility filter, 2005--2024. $\alpha_1+\beta_1$
near one is the usual high persistence; the standardised residuals are near unit
variance with heavy tails and (for equity) negative skew. LB$^2$ $p$ is the
Ljung--Box $p$-value on squared residuals at lag 10 ($>0.05$: volatility
clustering removed).}
\label{tab:garch}
\end{table}

\subsection{Surfaces}

\begin{figure}[H]
\centering
\includegraphics[width=\textwidth]{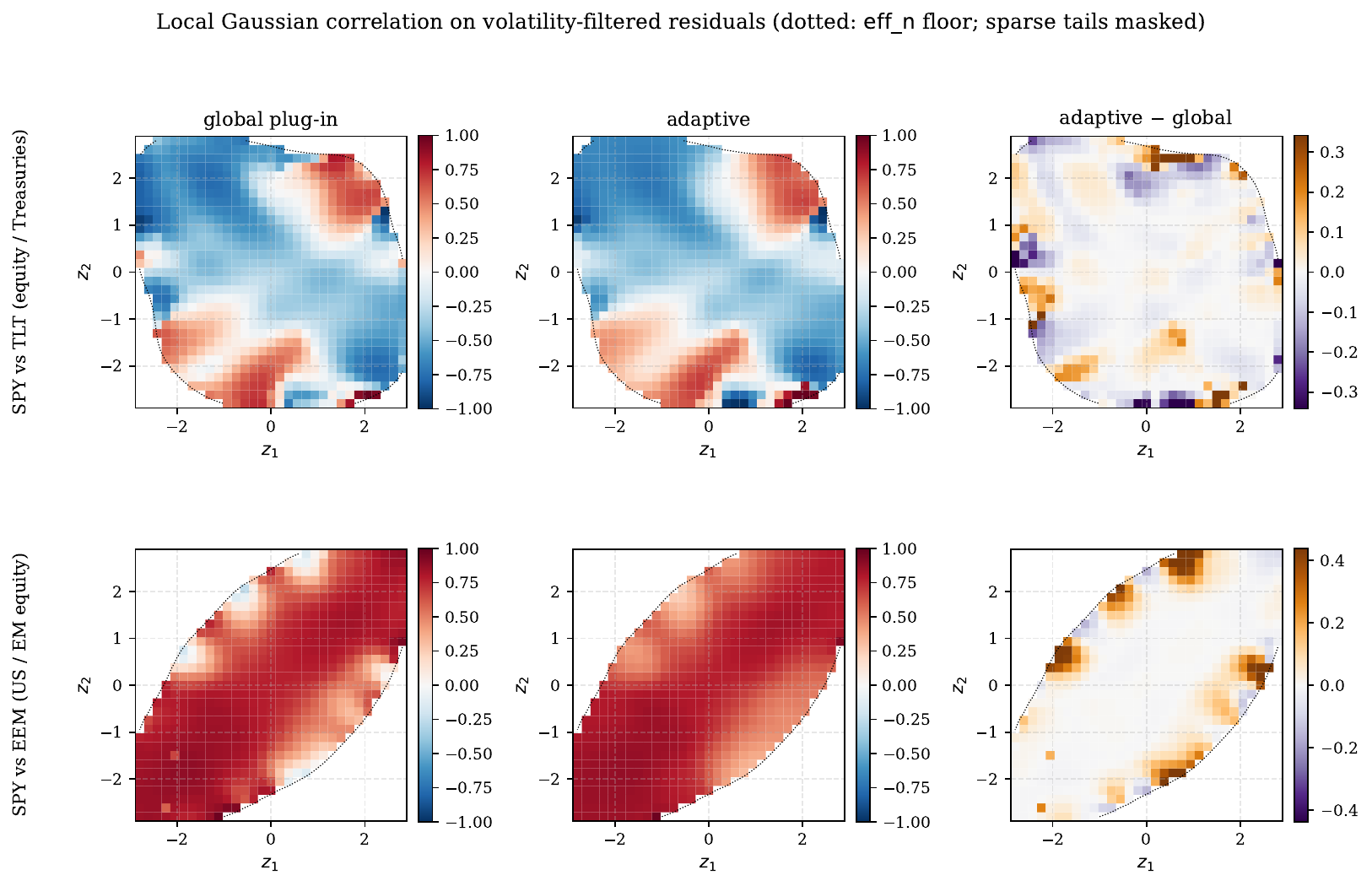}
\caption{Local Gaussian correlation on volatility-filtered residuals, global
plug-in (left) versus adaptive (centre) and their difference (right), for SPY/TLT
(top) and SPY/EEM (bottom). Dotted contour: the $\effn=15$ floor; the sparse
tails beyond it are masked.}
\label{fig:realsurf}
\end{figure}

Figure~\ref{fig:realsurf} shows the LGC surfaces, global versus adaptive, with
the sparse tails (local effective sample size below the Monte Carlo floor of
$15$) masked so the comparison is read only where either estimator has support.
The qualitative reading matches the simulation. For SPY/TLT the bulk dependence
is negative, with the local correlation strengthening and changing character
toward the corners; the adaptive estimator, spending its fixed smoothing budget
outward, presents the off-centre structure more coherently while the global
surface is granular there. For SPY/EEM the dependence is strongly positive
throughout, and the difference panel is almost one-signed in the tails: the
adaptive estimator resolves \emph{stronger} joint-tail correlation than the
global one, which over-smooths and speckles in the same sparse region. This is
the expected effect, redistribution toward the tails, on data where, unlike the
simulation, we cannot grade it against a truth.

\subsection{Stability under resampling}

\begin{figure}[H]
\centering
\includegraphics[width=\textwidth]{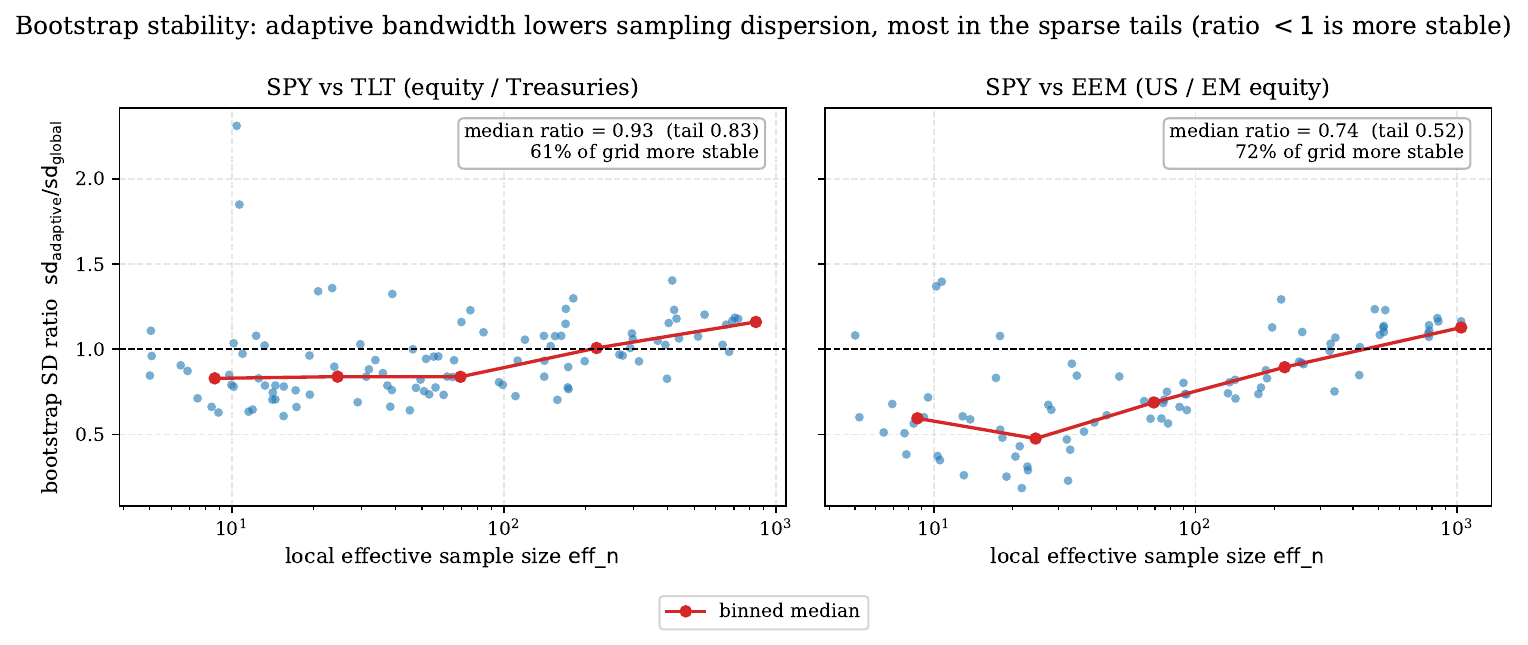}
\caption{Bootstrap SD ratio (adaptive/global) against local effective sample
size, by pair, with a binned median (red) and the parity line. Below one is more
stable; the gain is largest in the sparse tails and closes toward the dense
centre.}
\label{fig:realstab}
\end{figure}

Because the GARCH step pre-whitens to near-i.i.d.\ residuals, we resample the
residual pairs i.i.d.\ with replacement ($B=200$) and recompute both surfaces on
each resample. Figure~\ref{fig:realstab} plots the point-wise bootstrap standard
deviation ratio, adaptive over global, against local effective sample size. The
ratio sits below one across most of the grid and falls furthest in the sparse
tail, the regime the adaptive rule targets: median ratio $0.93$ for
SPY/TLT and $0.74$ for SPY/EEM, dropping to $0.83$ and $0.52$ respectively
in the tail band ($\effn<40$); the adaptive surface is more stable on $61\%$
and $72\%$ of the supported grid. The deep, near-empty corners are excluded
(both estimators are undefined there), so this is not a boundary artefact. The
simulation's stabilisation result thus carries over to real returns, and it does
so most where it matters, in the data-starved tails.

\begin{table}[H]
\centering\small
\begin{tabular}{llccc}
\toprule
Pair & $q$ & $\rho$ global & $\rho$ adaptive & $\effn$ \\
\midrule
\multirow{5}{*}{SPY/TLT}
 & 0.01 & $-0.32$ & $0.33$ & 7$^\dagger$ \\
 & 0.05 & $0.37$ & $0.33$ & 57 \\
 & 0.50 & $-0.29$ & $-0.28$ & 843 \\
 & 0.95 & $0.64$ & $0.63$ & 56 \\
 & 0.99 & $0.23$ & $0.55$ & 7$^\dagger$ \\
\midrule
\multirow{5}{*}{SPY/EEM}
 & 0.01 & $0.87$ & $0.86$ & 75 \\
 & 0.05 & $0.88$ & $0.87$ & 273 \\
 & 0.50 & $0.78$ & $0.79$ & 1035 \\
 & 0.95 & $0.83$ & $0.82$ & 258 \\
 & 0.99 & $0.81$ & $0.80$ & 61 \\
\bottomrule
\end{tabular}
\caption{Local correlation at the diagonal quantile points, global versus
adaptive, with the local effective sample size. For the positively-dependent
SPY/EEM pair the two estimates coincide in the well-populated diagonal and the
adaptive bootstrap standard errors (not shown) are smaller in the tails (e.g.\ at
$q=0.95$, $0.024$ vs $0.055$). $^\dagger$SPY/TLT is negatively dependent, so its
mass lies off the main diagonal and the extreme diagonal points are themselves
data-starved ($\effn\approx7$, below the floor of $15$); those two rows mark
scarcity rather than reliable estimates, and the surface (Fig.~\ref{fig:realsurf})
is the better view for that pair.}
\label{tab:realquant}
\end{table}

\subsection{What the application does and does not establish}

On real data the adaptive bandwidth produces tail-dependence surfaces that are
quantitatively more stable under resampling, the gain concentrated in the sparse
tails. What it cannot establish, here or anywhere without a known truth, is a
tail-\emph{accuracy} improvement. This caveat is sharper given the regime map of
\S\ref{sec:regime}: the SPY/EEM pair is strongly dependent, the very regime in which
adaptivity trades variance for bias, so its lower resampling dispersion may partly
reflect over-smoothing rather than better accuracy. We therefore claim only the
variance property (more stable surfaces) and not improved estimation. The deepest
corners, finally, remain beyond reach for either estimator, the same $\effn$ floor
seen in the Monte Carlo: the binding constraint is genuine scarcity, which no
bandwidth cures.

\section{Discussion}
\label{sec:discussion}

Our findings form one arc. The binding constraint on tail LGC is local scarcity, not
the marginal model; scarcity expresses itself as a Fisher variance floor; that floor
is the variance term of the local AMISE; and minimising the AMISE yields a
location-specific optimal bandwidth. But the same derivation explains why that
bandwidth is not a panacea: a location-varying bandwidth can only \emph{redistribute}
local sample size, and that helps the bias--variance trade only where the surface is
curved enough to make the trade worthwhile and populated enough to make it stable.
That band, moderate dependence, is narrow, and outside it a single global bandwidth
is preferable, sharply so at strong dependence. Our study turns the field's informal
experience (an adaptive bandwidth was tried and found inferior; global selection is
the default) into a quantitative account of \emph{why} and \emph{where}.

The contribution is therefore threefold: a diagnostic that identifies
the variance floor, the first AMISE-optimal local bandwidth for LGC together with its
validated bias expansion, and a map of the one regime in which location-adaptivity
pays. The boundaries are as informative as the band. On a flat surface (weak
dependence) redistribution only adds variance; on a steep surface (strong dependence)
it incurs bias that grows with the sample size; and in the deepest, near-empty corners
no bandwidth manufactures information the data do not contain. On real returns the
adaptive estimator is more stable under resampling, but, absent a ground truth and
in light of the strong-dependence bias, we read that as a variance rather than an
accuracy property, and report it as such (\S\ref{sec:real_data}).

The open problem that remains is a bandwidth robust \emph{across} dependence
strength. The failure mode is specific: the pointwise-AMSE allocation over-smooths
the steep mid-region of a strongly dependent surface, and the oracle-scale
experiment of \S\ref{sec:regime} shows that no choice of overall scale repairs
it; what must change is the shape. That points to an integrated- rather than
pointwise-AMISE criterion, or an $\effn$-aware bound
tied directly to the variance floor of \S\ref{sec:scarcity}. Two further extensions:
the exact closed-form constant in the bias functional~\eqref{eq:beta}, a full
specialization of \citet{hjort1996locally} to the bivariate Gaussian family, would
replace the plug-in $\beta$ with an analytic one; and the same programme extends to
the multivariate locally Gaussian density estimator of \citet{otneim2017locally}.

\section*{Code and data availability}
All code, simulation harnesses, and figure scripts that reproduce the results are
openly available at \url{https://github.com/akashdeepo/LGC_NTS_Pilot}. The Monte
Carlo uses deterministic cross-process seeding, so every reported number is
reproducible from the repository; the real-data application fetches public daily
prices and caches the volatility-filtered residuals used here.

\section*{Acknowledgements}
The authors thank Prof.\ B\aa{}rd \Stove\ for suggesting the adaptive-bandwidth
direction and pointing to \citet{brockmann1993locally}, which prompted the
development reported here.

\bibliographystyle{plainnat}
\bibliography{references}

\end{document}